\documentclass[reprint, amsmath, amssymb, 
aps,prb,superscriptaddress,longbibliography,floatfix]{revtex4-1}
\usepackage{graphicx}% Include figure files
\usepackage{color}
\usepackage{bm}% bold math
\usepackage[breaklinks=true,colorlinks,citecolor=blue,linkcolor=blue,urlcolor=blue]{hyperref}
\def\be {\begin{equation}}
\def\ee {\end{equation}}
\def\bea{\begin{eqnarray}}
\def\eea{\end{eqnarray}}
\def\nn{\nonumber}

\begin{document}
\title{Collective plasmonic modes in the chiral multifold fermionic material CoSi} 
\author{Debasis Dutta}
\thanks{These authors contributed equally to this work}
\affiliation{Department of Physics, Indian Institute of Technology Kanpur, Kanpur 208016}
\author{Barun Ghosh $^*$}
\email{barunghosh02@gmail.com}
%\thanks{These authors contributed equally to this work}
\affiliation{Department of Physics, Northeastern University, Boston, Massachusetts 02115, USA}
\author{Bahadur Singh}
\affiliation{Department of Condensed Matter Physics and Materials Science,Tata Institute of Fundamental Research, Mumbai 400005, India}
\author{Hsin Lin}
\affiliation{Institute of Physics, Academia Sinica, Taipei 11529, Taiwan}
\author{Antonio Politano}
\affiliation{INSTM and Department of Physical and Chemical Sciences, University of L'Aquila, via Vetoio 67100 L'Aquila (AQ), Italy; CNR-IMM Istituto per la Microelettronica e Microsistemi, I-95121 Catania, Italy;}
\author{Arun Bansil}
\affiliation{Department of Physics, Northeastern University, Boston, Massachusetts 02115, USA}
\author{Amit Agarwal}
\email{amitag@iitk.ac.in}
\affiliation{Department of Physics, Indian Institute of Technology Kanpur, Kanpur 208016}

\begin{abstract}
Plasmonics in topological semimetals offers exciting opportunities for fundamental physics exploration as well as for technological applications. Here, we investigate plasmons in the exemplar chiral crystal CoSi, which hosts a variety of multifold fermionic excitations. 
We show that CoSi hosts two distinct plasmon modes in the infrared regime at 0.1 eV and 1.1 eV in the long-wavelength limit. The 0.1 eV plasmon is found to be highly dispersive, and originates from intraband collective oscillations associated with a double spin-1 excitation, while the 1.1 eV plasmon is dispersionless and it involves interband correlations. Both plasmon modes lie outside the particle-hole continuum and possess long lifetime. Our study indicates that the CoSi class of materials will provide an interesting materials platform for exploring fundamental and technological aspects of topological plasmonics.
\end{abstract}
\maketitle

\section{Introduction}
Topological semimetals have emerged as an exciting class of materials for hosting a variety of new types of quasi-particle excitations ~\cite{Bansil16,DSM_WSM_review_viswanath,Na3Bi_PRB,
Cd3As2_PRB,Z2NL_Fu,Fang_2016,Wang17,hourglass_Ag2BiO3,PhysRevX.6.031003}. Energy dispersions of charge carriers in topological semimetals inherently differ from those in ordinary metals and, therefore, we would expect fundamental differences in their collective excitations\cite{Pines1952,Pines1962,giuliani2005quantum,SDSharma2009,def_fo_plas_WSM,RashiSachdeva2015,Hofmann2015,Thakur_2017,PhysRevB.97.035403,Sadhukhan2020,Wang2021}. Moreover, the relatively low electron density in a topological semimetal makes it possible to drive plasmonic excitations in the terahertz and infrared regimes with potential applications in biochemical sensing, medical treatment, solar cells, and  optoelectronics, among other areas~\cite{StefanoLupi2020,Maier2007,Antonio2017,C8NR01395K}. 

A condensed matter system can host elementary excitations that do not have any high-energy counterparts. Examples include, type-II Dirac/Weyl, and threefold, fourfold, sixfold, and eightfold degenerate unconventional fermions~\cite{Soluyanov2015,Yan2017,PhysRevX.6.031003,Bradlynaaf5037}. Plasmons in topological insulators, and Dirac, Weyl and nodal-line semimetals are attracting substantial current interest. Experimental observations include the Dirac plasmons in graphene and in type-II Dirac semimetal PtTe$_2$~\citep{Grigorenko2012, DiPietro2013,PtTe2PRL,BGhosh2020,Wang2019}. The field of topological plasmonics, however, remains largely unexplored. 

\begin{figure}[t!]
	\includegraphics[width =0.9\linewidth]{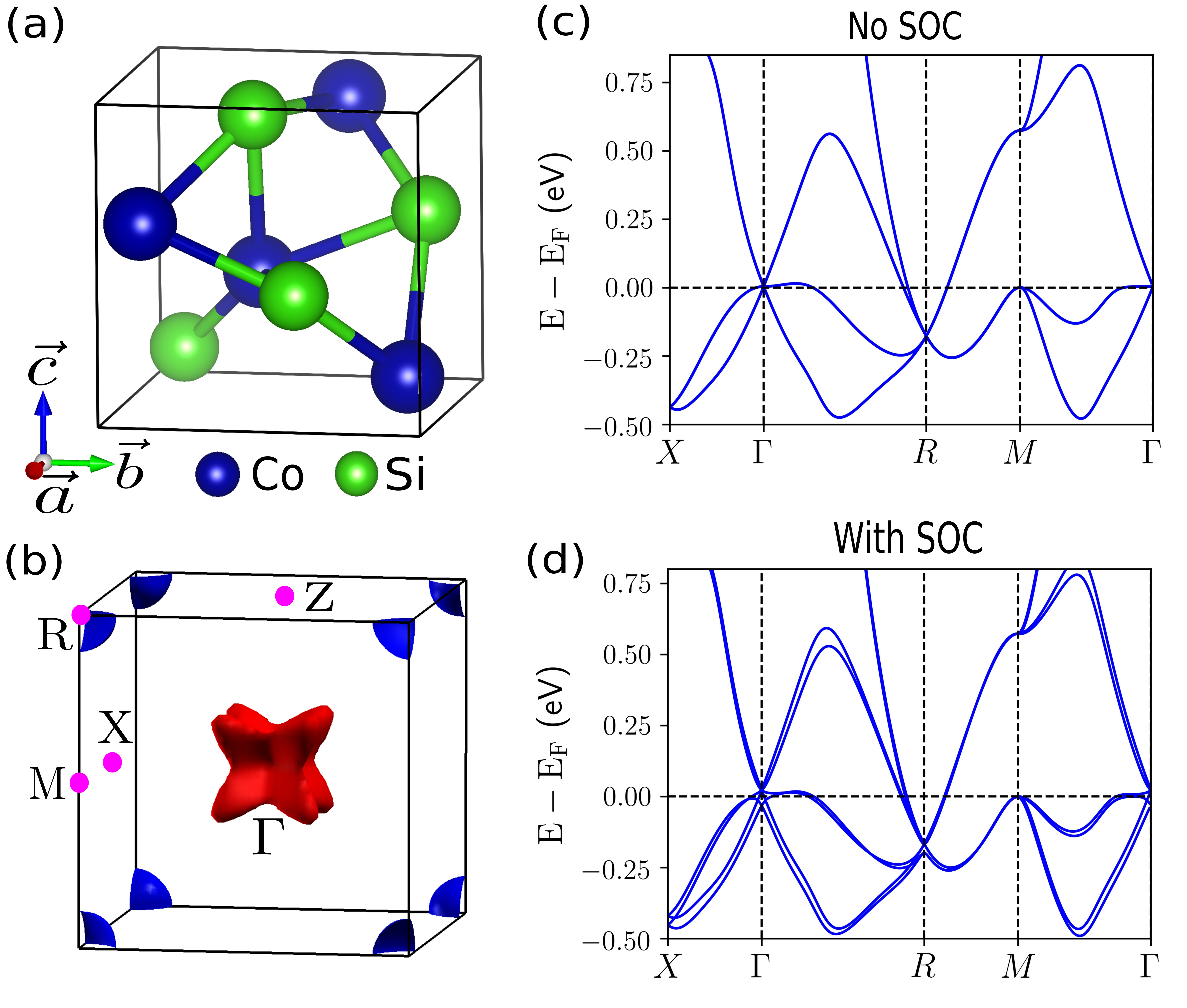}	
	\caption{ (a) Crystal structure of CoSi. (b) Fermi surface of CoSi in the cubic Brillouin zone (BZ). The red and blue colors denote the hole and electron pockets, respectively. The electronic band structure of CoSi (c) before and (d) after including the spin-orbit coupling. }
	\label{fig1}
\end{figure} 

The transition-metal silicide family of chiral compounds, including CoSi, RhSi etc., has attracted a lot of recent interest~\cite{ShouCheng2017,RhSi,Reeseaba0509}. This is predominantly inspired from the structural chirality in these topological materials which leads to various types of chiral multifold fermionic excitations. 
%Specifically, in the absence of spin-orbit coupling,  CoSi hosts a threefold spin-1 excitation at the zone center ($\Gamma$), which is transformed into a fourfold spin-3/2 Rarita-Schwinger-Weyl (RSW) fermion and a twofold degenerate spin-1/2 Weyl fermion by the spin-orbit coupling (SOC). 
Specifically, in the absence of spin-orbit coupling,  CoSi hosts a threefold (for each spin channel) spin-1 excitation at the zone center ($\Gamma$). In presence of SOC, it is transformed into a fourfold spin-3/2 Rarita-Schwinger-Weyl (RSW) fermion and a twofold degenerate spin-1/2 Weyl fermion.
At the zone boundary (R-point), CoSi hosts a double-Weyl fermion, which also transforms into a sixfold degenerate double spin-1 excitation in the presence of SOC. These chiral fermions carry an integer topological charge. In fact, the CoSi family of materials hosts a chiral charge of $\pm 4$, which is the highest possible chiral charge in a non-magnetic chiral topological semimetal. This leads to large Fermi arcs in the surface states, as confirmed by recent angle resolved photoemission spectroscopy (ARPES) and quasi particle interference (QPI) experiments~\citep{PhysRevLett.122.076402,Rao2019,Yuaneaaw9485}. The bulk charge carriers have also been predicted to show interesting optical properties including large circular photogalvanic response, which has been recently verified experimentally ~\cite{Ni2021}. However, the collective density excitation or plasmons in the CoSi family remain unexplored. 

In this paper, using low-energy models and first-principles calculations, we unveil the dielectric and plasmonic properties of CoSi family of materials. We show that the low energy collective excitations in CoSi primarily consist of two plasmon modes. The lowest energy mode, which originates from the collective mode of the spin-1 excitation, has a plasmon gap of 0.1 eV, and is highly dispersive in nature. The higher energy mode with a plasmon gap of 1.1 eV is virtually dispersionless and its origin is tied to the interband correlations. Both these plasmon modes live outside the intraband particle-hole continuum and are long-lived.

The remainder of this paper is organized as follows:  In Sec.~\ref{sec2}, we discuss the crystal structure and the electronic properties of the CoSi family of compounds. This is followed by a discussion on the low-energy description of various chiral multifold fermions and their collective modes in Sec.~\ref{sec3}.  Section~\ref{sec4} is dedicated to the {\it ab-initio} results of the dielectric function and the low energy plasmon modes in CoSi.  In Sec.~\ref{sec5} we discuss the effect of doping (or tuning the Fermi level) on the low energy plasmon modes in CoSi and summarize our findings in Sec. ~\ref{sec6}.
\section{crystal structure and Electronic properties of $\ {\bf CoSi}$}
\label{sec2}
CoSi belongs to the transition metal silicide family of compounds which crystallize in the chiral non-symmorphic space group P$2_13$ (group number 198) with a cubic unit cell. The unit cell contains four formula units of CoSi with the Co and Si atoms both occupying the 4a-Wycoff site [see Fig.~\ref{fig1}(a)]. The space group contains twelve symmetry operations that can be generated by a threefold rotation symmetry along the (111) direction and twofold screw symmetries along the $z-$ and $x$-axis. It lacks any inversion, mirror, and roto-inversion symmetries. 
\begin{figure*}[t!]
	\includegraphics[width = 0.92\textwidth]{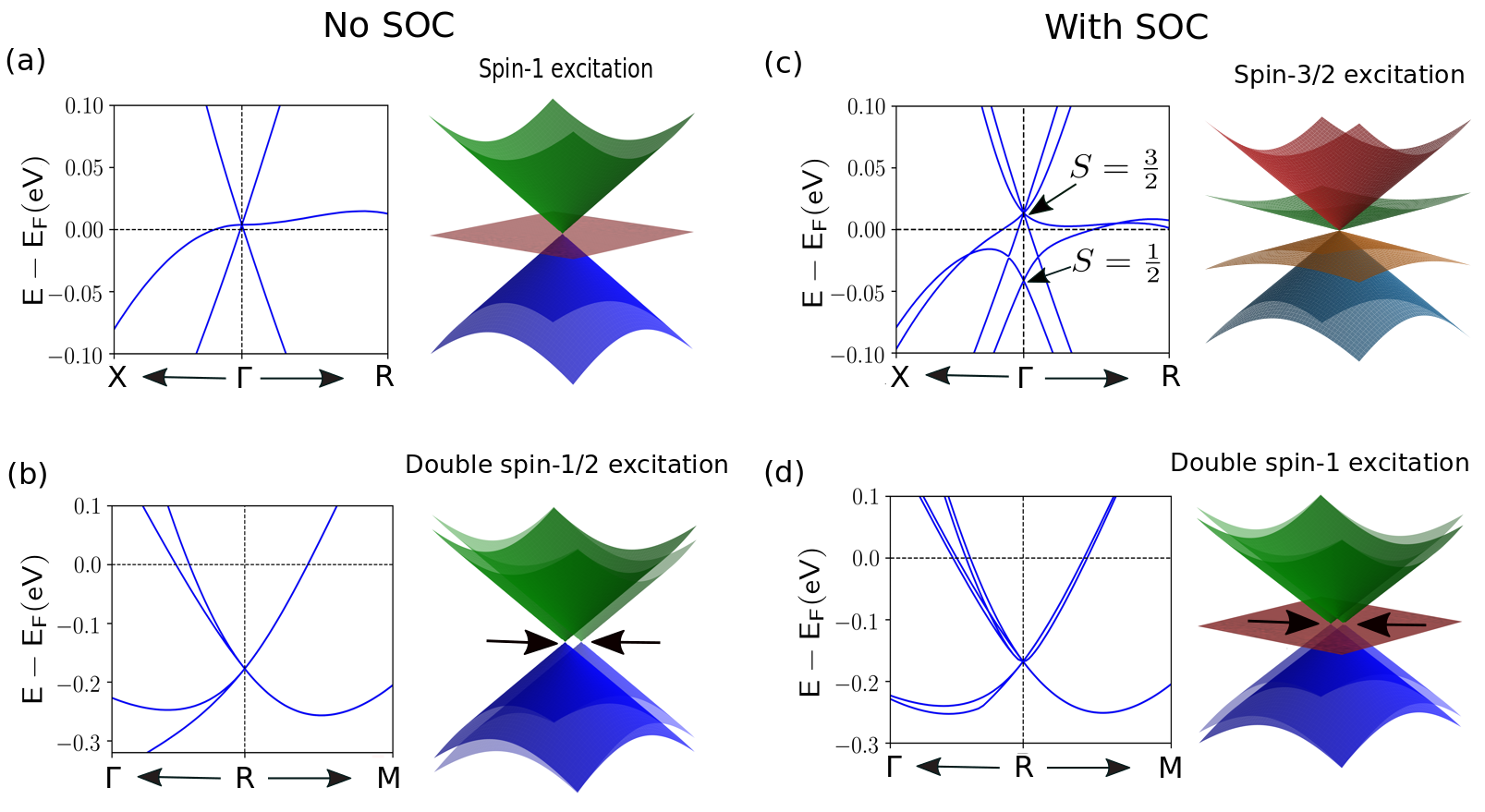}	
	\caption{Schematic illustration of the dispersion along with the DFT calculated band structure for various multifold fermions in CoSi. In the absence of SOC, there is (a) spin-1 excitation around the $\Gamma$-point and (b) double spin-1/2 excitation around the $R$ point. After the effects of SOC are included, these are transformed to (c) spin-3/2 fermionic excitation around the $\Gamma$ point and (d) double spin-1 excitation around the $R$ point.}
	\label{fig2}
\end{figure*} 

There are only four bands near the Fermi level for each spin channel in the absence of SOC which give rise to different low energy excitations in CoSi [see Fig.~\ref{fig1}(c)]~\cite{ShouCheng2017}. These bands mostly originate from the $d-$orbitals of Co and the $p-$orbitals of Si. There is a threefold degenerate band crossing for each spin channel at the $\Gamma$ point and a fourfold degenerate band crossing at the $R$ point of the BZ. The threefold band crossing at the $\Gamma$ point has a topological charge of $C=2$ and represents a pseudospin-1 excitation. At the same time, the fourfold crossing point at the $R$ point also carries a topological charge of $C= -2$, and it hosts the double Weyl fermion [see Fig.~\ref{fig2}(a)-(b)].

As we include the SOC effect, all bands are split except at the time-reversal invariant momentum points due to the absence of the inversion symmetry [see Fig.~\ref{fig1}(d)]. The six-fold degeneracy (including both spin channels) at the band crossing located at the $\Gamma$ point reduces to a fourfold degenerate band crossing, separated in energy from a twofold degenerate band crossing with both crossings lying in the vicinity of the Fermi energy. Similarly, the eightfold degeneracy at the $R$  point is reduced to a six-fold crossing and a twofold degenerate crossing. The fermionic quasiparticle excitation at the $\Gamma$ point represent a massless spin-3/2 fermion and a spin-1/2 fermion, while that at the $R$ point represents a double spin-1 excitation which is stabilized by the time-reversal symmetry [see Fig.~\ref{fig2}(c),(d)]. The topological charges associated with the carriers in the vicinity of the band crossing at the $\Gamma$ and $R$  point is $C=\pm 4$.

The low energy bands around the Fermi level give rise to distinct electron and hole pockets in the BZ as shown in Fig.~\ref{fig1}(b). The pocket around the $R$  point is an electron pocket, while the one around the $\Gamma$ point is a hole pocket. These electron and hole pockets should give rise to distinct collective density excitations, corresponding to different multi-fold chiral fermions.

\section{Chiral multifold fermions and plasmons}
\label{sec3}
Multifold chiral fermions are the generalization of the twofold degenerate Weyl fermionic excitations with three-, four- and sixfold degeneracy. Similar to the Dirac and Weyl quasiparticle excitations in crystals, these are also protected by crystal symmetries. The low energy excitations around these symmetry-protected multi-fold band crossings can be described by a generalization of Weyl Hamiltonian with an arbitrary spin. The generic Hamiltonian is of the form ${\cal H}=\hbar v_F \bm{k} \cdot \bm{S}$, where $\bm{k}$ is the crystal momentum and the matrices denoting the components of ${\bm S} = \{S_x,S_y,S_z\}$ take the form of higher spin representation of $SU(2)$~\cite{Flicker2018}.

\subsection{Multi-fold fermions}
The lowest degeneracy (two-fold) multifold fermion, {\it i.e}, Weyl fermion is described by the Weyl Hamiltonian ${\cal H}_{2f}=\hbar v_F\bm{k}\cdot\bm{\sigma}$, where $\sigma$ is Pauli spin vector for spin-1/2 fermions. Going from spin-1/2 to spin-1 excitation, the threefold Hamiltonian takes the form ${\cal H}_{3f}({\bm k},\pi/2)=\hbar{v_F}{\bm k}\cdot\bm{S}_1$, where the components of ${\bm S}_1$ form a spin-1 representation of SU(2). In vicinity of the threefold degenerate point, the most general $\bm{k} \cdot \bm{p}$ Hamiltonian takes the form
\begin{equation}
	{\cal H}_{3f}({\bm k},\phi)=\hbar{v_F}\begin{pmatrix} 0 & e^{i\phi}k_x & e^{-i\phi}k_y \\
	 e^{-i\phi}k_x & 0 & e^{i\phi}k_z \\
	 e^{i\phi}k_y & e^{-i\phi}k_z & 0 \\
	\end{pmatrix}~.
	\label{equ_spin1}
\end{equation}  
Here, $v_F$ is the Fermi velocity and $\phi$ is a material dependent parameter~\cite{Bradlynaaf5037,Flicker2018}.  In the absence of SOC, time-reversal symmetry restricts $\phi=\pi/2$.  In this case, the band energies for the threefold fermions are $E_s=s\hbar{v_F}|{\bm k}|$, with $s=-1,0,1$, similar to the expectation value of the $z-$spin component for a spin-1 system. In CoSi without SOC, the threefold degenerate band crossing at the $\Gamma$ point (for each spin channel) is described by this spin-1 Weyl Hamiltonian [see Fig.~\ref{fig2}(a)]. 

In the absence of SOC the fourfold degenerate band crossing at the $R$-point can be described by a double (spin-1/2) Weyl fermion for each spin channel. The low energy Hamiltonian of this fourfold degenerate double Weyl fermion comprises of two decoupled copies of single Weyl nodes at the same momentum~\cite{deJuan2019} [see Fig.~\ref{fig2}(b)]. The model Hamiltonian is expressed as,
\begin{equation}
	{\cal H}_{\rm DW}({\bm k})=\hbar{v_F}\begin{pmatrix} \bm{k}\cdot\bm{\sigma} & 0 \\
	0 & \bm{k}\cdot\bm{\sigma}
	\end{pmatrix}~.
	\label{DoubleWeyl}
\end{equation}
Since each of such fourfold degenerate points is composed of two Weyl cones of the same chirality, the total charge associated with such band crossings is $C = \pm 2$. 

An alternate way to have fourfold degenerate fermions is via a Hamiltonian of the form ${\cal H}_{4f}({\bm k})=\hbar{v_F}\boldsymbol{k}\cdot\boldsymbol{S}_{3/2}$, where $\boldsymbol{S}_{3/2}$ are three matrices that form a spin-3/2 representation of $SU(2)$. These fourfold fermions are found in materials belonging to tetrahedral or octahedral groups, including CoSi, in the presence of SOC.
A more general ${\bm k}\cdot{\bm p}$ Hamiltonian for the fourfold fermion in materials of the octahedral group is given by~\cite{Flicker2018,deJuan2019}
\begin{equation}
{\cal H}_{4f}= \begin{pmatrix} ak_z & 0 & -\frac{a+3b}{4}k_{+} & \frac{\sqrt{3}(a-b)}{4}k_{-} \\
0 & bk_z & \frac{\sqrt{3}(a-b)}{4}k_{-} & -\frac{3a+b}{4}k_{+}  \\ 
-\frac{a+3b}{4}k_{-} & \frac{\sqrt{3}(a-b)}{4}k_{+} & -ak_z &0 \\
\frac{\sqrt{3}(a-b)}{4}k_{+} &-\frac{3a+b}{4}k_{-} &  0 & -bk_z\\ 
\end{pmatrix}~.
\label{eq4f}
\end{equation}
Here, $k_{\pm}=k_x\pm ik_y$, and ${a}=\hbar{v}_F\cos\chi$ and ${b}=\hbar{v}_F\sin\chi$, and $\chi$ is a material dependent parameter. 
The Hamiltonian in Eq.~\eqref{eq4f} recovers full rotational symmetry when $\chi=\rm{arctan}(-3)$ or $\chi=\rm{arctan}(-1/3)$ and it takes the form ${\cal H}_{4f}({\bm k})=\hbar{v_F}\boldsymbol{k}\cdot\boldsymbol{S}_{3/2}$.  In this case, the energy eigenvalues are given by $E_s=2s\hbar{v_F}|{\bm k}|$, with $s=-\frac{3}{2},-\frac{1}{2},~\frac{1}{2}$, and $\frac{3}{2}$. On switching the SOC in CoSi, the sixfold degenerate bands (including both spin channel) at the $\Gamma$ point get split in a band crossing of fourfold degenerate spin-3/2 fermion and an additional doubly degenerate band crossing of spin-1/2 Weyl fermion [see Fig.~\ref{fig2}(c)].

In addition to these, there is another sixfold degenerate excitation found at the $R$-point in CoSi with SOC [see Fig.~\ref{fig2}(d)]. This is a `double spin-1' excitation with sixfold degeneracy that is also protected by cubic symmetry.  Similar to the double Weyl fermion, this comprises of two decoupled copies of the spin-1 Weyl Hamiltonian specified in Eq.~\ref{equ_spin1}.
The low energy Hamiltonian of such sixfold fermions, to linear order in ${\bm k}$, is given by~\cite{Flicker2018} 
\begin{equation}
	{\cal H}_{6f}({\bm k})=\begin{pmatrix}
	{\cal H}_{3f}({\bm k},\frac{\pi}{2}-\phi) & 0\\
	0 & {\cal H}_{3f}({\bm k},\frac{\pi}{2}+\phi)
	\end{pmatrix}~.
	\label{Doublespin1_H}
\end{equation}
Similar to the case of spin-1 excitation, $\phi$ is set to $\pi/2$ in presence of time-reversal symmetry. 

Putting everything together, we find that in the presence of SOC, CoSi hosts i) fourfold spin-3/2 fermions at the $\Gamma$ point, ii) spin-1/2 Weyl fermions at the $\Gamma$ point, and iii) sixfold degenerate double spin-1 fermions at the $R$ point.

\subsection{Plasmon modes in multifold fermions}
\begin{table*}[ht!]
	\caption{Long wavelength plasmon energy gap ($\hbar\omega_p$) for different linearly dispersing multifold fermions in three dimensions. 
	The weak Coulomb interaction limit is specified by $\kappa > e^2/(\hbar v_F)$ or $\alpha_{{\rm fine}} < 1$. 
	For simplicity of notation, we have defined $\Delta_p = \mu \sqrt{2 \alpha_{\rm fine}/(3 \pi)}$.
	}
	\begin{tabular}{c c c c  }
		\hline \hline 
		Type & Hamiltonian & Plasmon gap: $\hbar\omega_p=$ & Weak interaction: $\hbar\omega_p = $ \\ 
		\hline 
		Twofold spin-1/2 (Weyl) & ${\bm k}\cdot\boldsymbol{\sigma}$ & $ \Delta_p \left[ 1+ \frac{\alpha_{{\rm fine}}}{6\pi}\ln\left|\frac{\hbar^2\Lambda^2}{4\mu^2-\hbar^2\omega_p^2} \right| \right]^{-1/2}$ & $ \Delta_p$  \\ 
		\hline
			Threefold spin-1 & ${\bm k}\cdot\boldsymbol{S}_1$ & $ \Delta_p \left[ 1+ \frac{2\alpha_{{\rm fine}}}{3\pi}\ln\left|\frac{\hbar^2\Lambda^2}{\mu^2-\hbar^2\omega_p^2} \right| \right]^{-1/2}$  & $\Delta_p$  \\
		\hline
		Fourfold double spin-1/2 & $\begin{pmatrix}
		{\bm k}\cdot\boldsymbol{\sigma} & 0 \\
		0 & {\bm k}\cdot\boldsymbol{\sigma}
		\end{pmatrix}$ & $ \sqrt{2}\Delta_p \left[ 1+ \frac{2\alpha_{{\rm fine}}}{6\pi}\ln\left|\frac{\hbar^2\Lambda^2}{4\mu^2-\hbar^2\omega_p^2} \right| \right]^{-1/2}$  & $\sqrt{2}\Delta_p$ \\					
		\hline
	
		Fourfold spin-3/2 & ${\bm k}\cdot\boldsymbol{S}_{3/2}$ & $ \frac{2}{\sqrt{3}}\Delta_p \left[ 1 + \frac{\alpha_{{\rm fine}}}{6\pi}\left( 4~\ln\left|\frac{\hbar^2\Lambda^2}{4\mu^2-\hbar^2\omega_p^2}\right| +3~\ln\left|\frac{36\mu^2-9\hbar^2\omega_p^2}{4\mu^2-9\hbar^2\omega_p^2}\right|   \right) \right]^{-1/2}$ & $\frac{2}{\sqrt{3}}\Delta_p$   \\ 
		\hline 
		Sixfold double spin-1 & $\begin{pmatrix}
		{\bm k}\cdot\boldsymbol{S}_1 & 0 \\
		0 & {\bm k}\cdot\boldsymbol{S}_1
		\end{pmatrix}$& $\sqrt{2}\Delta_p \left[ 1+ \frac{4\alpha_{{\rm fine}}}{3\pi}\ln\left|\frac{\hbar^2\Lambda^2}{\mu^2-\hbar^2\omega_p^2} \right| \right]^{-1/2}$   & $\sqrt{2}\Delta_p$   \\
		\hline  \hline
	\end{tabular} 
	\label{table:1}
\end{table*}

Given the electron and hole pockets associated with these different multifold fermions, a natural question to ask is what is the dispersion of the collective excitations in these multifold fermionic systems? Can they be observed in experiments on real materials hosting these? In this subsection, we focus on the first question, while the second question is addressed in the next section. 

To obtain the low energy dispersion of the plasmon modes, we calculate the interacting density-density response function and the dielectric function within the random phase approximation(RPA)~\cite{Pines1962,giuliani2005quantum}. Within RPA, the wave-vector (${\bm q}$) and frequency ($\omega$) dependent longitudinal dielectric function is given by
\begin{equation}
\epsilon^{\rm RPA}({\bm q},\omega)=1-V(q)\chi^{\rm NI}({\bm q},\omega)~.
\label{epsiloneq}
\end{equation}
Here, $V(q)\equiv\frac{4\pi e^2}{\kappa q^2}$ is the Fourier transform of the Coulomb interaction in three dimensions, with $\kappa$ being the effective background dielectric constant. In Eq.~\eqref{epsiloneq}, $\chi^{\rm NI}(q,\omega)$ is the non-interacting density-density response function [see Eq.~\eqref{chi0} in Appendix \ref{chi0_calculation} for more details]. Within RPA, the plasmon dispersion can be obtained from roots of the complex dielectric function, 
\begin{equation}
	\epsilon^{\rm RPA}(q,\omega)=0~.
	\label{root_epsilon}
\end{equation}
For a given ${q}$, the complex root of Eq.~\eqref{root_epsilon} is specified by $\omega=\omega_p -i\Gamma$. Here $\Gamma_p$ is related to the damping rate of the plasmon mode at frequency $\omega_p$, which mainly arises from the single particle excitations or finite imaginary part of dielectric function~\cite{giuliani2005quantum,Agarwal2015,Agarwal2015B}.  

In three dimensions, the plasmon dispersion is typically gapped, and for small $q$ it has the form of $\omega(q) = \omega_p + a q^2 + {\cal O}(q^4)$. Here $\omega_p$ is the plasmon gap in the long wavelength limit, and $a$ is a material dependent parameter. 
Focusing on the plasmon gap corresponding to different multifold fermions, we work in the {\it dynamical} long wavelength limit ($q \to 0$ and $\omega\gg {v_Fq}$). In this limit, at the plasmon frequency ${\rm Im}[\epsilon^{\rm RPA}(q\to{0},\omega)]=0$, and Eq.~\eqref{root_epsilon} reduces to
\begin{equation}
	1- V(q){\rm Re}\left[\chi^{\rm NI}(q\to{0},\omega)\right]=0~.
	\label{root_epsilon1}
\end{equation}
The details of the calculation of the long wavelength limit of $\chi^{\rm NI}$ for different multifold fermion Hamiltonian are presented in Appendix~\ref{chi0_calculation}. For the case of a twofold degenerate single Weyl node, having a finite chemical potential $\mu$ (measured from the Weyl point) in the conduction band, the real part of the dielectric function can be expressed as
\begin{equation}
 {\rm Re}\left[\epsilon^{\rm RPA}(q\to{0},\omega)\right]= 1- V(q) {\rm Re}[\chi^{\rm NI}_{\rm interband} + \chi^{\rm NI}_{\rm intraband}]~.
\end{equation}
This can be evaluated as, 
\begin{equation}
 {\rm Re}\left[\epsilon^{\rm RPA}\right] =  1 + \frac{\alpha_{\rm {\rm fine}}}{6\pi}\ln\left|\frac{\hbar^2\Lambda^2}{4\mu^2-\hbar^2\omega^2} \right| - \frac{2\alpha_{\rm {\rm fine}}}{3\pi} \frac{\mu^2}{\omega^2}.
 \label{Reepsilon}
\end{equation}
Here, $\alpha_{\rm {\rm fine}}\equiv{e^2}/(\kappa\hbar{v_F})$ is the effective fine structure constant related to the strength of Coulomb interaction, and we have introduced a high energy cutoff ($\equiv \hbar\Lambda$) for the linearly dispersing bands.  Using Eq.~\eqref{root_epsilon1}, the plasmon gap $\omega_p$ is given by the solution of the transcendental equation, 
\begin{equation}
	\hbar\omega_p= \Delta_p \left[1+ \frac{\alpha_{\rm fine}}{6\pi}\ln\left|\frac{\hbar^2\Lambda^2}{4\mu^2-\hbar^2\omega_p^2} \right| \right]^{-1/2}~.
	\label{weyl_plasmon}
\end{equation}
Here, we have defined 
\begin{equation}
	\Delta_p \equiv \mu \sqrt{\frac{2\alpha_{{\rm fine}}}{3\pi}}~,
\end{equation}
which is the plasmon gap in the weak interaction limit of $\kappa \gg e^2/{\hbar v_F}$ or $\alpha_{{\rm fine}}\ll 1$. For strong interactions, Eq.~\eqref{weyl_plasmon} can be interpreted in terms of the fine structure developing dynamical corrections arising from the presence of the infinite Fermi sea~\cite{def_fo_plas_WSM,Zhou2015}. 

We perform a similar calculation for all the known linearly dispersing multifold fermions. The calculated condition for the plasmon gap and the plasmon gap in the weak interaction limit is tabulated in Table~\ref{table:1}. For all linearly dispersing multifold fermions, we find that the plasmon gap $\Delta_p \propto \mu$, which implies $\Delta_p \propto n^{1/3}$, where $n$ is the free carrier density. For a given Fermi velocity and chemical potential, the plasmon gap is enhanced for a fourfold double spin-1/2 fermion, fourfold spin-3/2 fermions, and sixfold double spin-1 excitation compared to that of a spin-1/2 Weyl fermion. In contrast, it remains the same for a spin-1 excitation.
 
We now focus on material specific calculation in CoSi, which hosts these multifold fermionic quasiparticles. In the real material case, we only show the results including the effect of SOC.

\section{Plasmons in ${\bf CoSi}$}
\label{sec4}
\subsection{Methodology and computational details}\label{Methodology}
To understand the electronic band structure of CoSi, we perform first-principles calculations within the framework of DFT using the Vienna $Ab~initio$ Simulation Package~(VASP)~\cite{Kresse1996}. The projector augmented wave (PAW)  pseudopotentials are used within the generalized gradient approximation (GGA) scheme developed by Perdew-Burke-Ernzerhof (PBE) ~\cite{Kresse1999,Perdew1996}. We use an energy cutoff of $450$ eV for the plane-wave basis set and a $15\times{15}\times{15}$ Monkhrost-Pack k-grid for the BZ integration needed for the self-consistent calculation~\cite{mp_kgrid}. The response function calculations are performed with a much denser $k$-grid of $120\times 120 \times 60$. 

To calculate the low energy plasmon modes in CoSi, we start with the DFT calculated band structure. It is then used to construct the Wannier function based tight binding model using the dominant orbitals (Co $d$ and Si $p$ orbitals) in the vicinity of the Fermi energy~\cite{Marzari2012,Mostofi2014}. Using the obtained tight-binding model parameters, we calculate the dynamical form factor\cite{Thygesen2011} defined as 
%\begin{eqnarray}
%\label{form_fac}
%S^{\rm NI}({\bm q},\omega)= & & \frac{g}{\Omega}\sum_{{\bm k},n,n^{\prime}}(f_{n{\bm k}}-f_{n^{\prime}{\bm k}+{\bm q}})|{\langle{ {\bm{k+q},n^{\prime}}|}{{\bm k},n}\rangle}|^2 \nonumber \\
%& & \times\delta(\hbar\omega+\epsilon_{n{\bm k}} - \epsilon_{n^{\prime}{\bm k}+{\bf q}})~.
%%e^{i{\bm q}.\textbf{r}}|
%\end{eqnarray}
%%%%%%%%%%%%%%%%%%%%%%%%%%
\begin{eqnarray}
\label{form_fac}
S^{\rm NI}({\bm q},\omega)= & & \frac{g}{\Omega}\sum_{{\bm k},n,n^{\prime}}(f_{n{\bm k}}-f_{n^{\prime}{\bm k}+{\bm q}})F_{nn^{\prime}}(\bm{k},\bm{k+q}) \nonumber \\
& & \times\delta(\hbar\omega+\epsilon_{n{\bm k}} - \epsilon_{n^{\prime}{\bm k}+{\bf q}})~.
\end{eqnarray}
%Here, $\hbar{{\bm q}}$ and $\hbar{\omega}$ denote the momentum and energy transfer of the excitation and $\epsilon_{n{\bm k}}$ is the energy for the ${\bm k}$ state of the $n$th band - $|{\bm k},n\rangle$. 
Here,  $F_{nn^{\prime}}(\bm{k},\bm{k+q})=|\langle \psi_{n\bm{k}}|e^{-i\bm{q}\cdot\bm{r}}|\psi_{n^{\prime}\bm{k+q}}\rangle|^2=|\langle u_{n\bm{k}}|u_{n^{\prime}\bm{k+q}}\rangle |^2$ represents the absolute value of the square of the density fluctuation operator for band pair $(n,n')$ at the wave vector {\bf {q}} ~\cite{Kresse2006B,cohen_louie_2016,Wang2021,Giustino2016}.

In Eq.~\eqref{form_fac}, $\epsilon_{n{\bm k}}$ is the eigenvalues of the Bloch state $|\psi_{n\bm k}\rangle$ for band index $n$ and wavevector ${\bm k}$, $|u_{n\bm{k}}\rangle$ represents the cell periodic part of $|\psi_{n\bm k}\rangle$, $g$ denotes the spin degeneracy factor, and $\Omega$ is the unit cell volume. The summation on ${\bm k}$ runs over the first BZ and the band occupancy is given by the Fermi function ($f_{n{\bm k}}$). For simplicity, we work at zero temperature for which the Fermi function is given by a step function, $f_{n{\bm k}}=\theta(\mu-\epsilon_{n{\bm k}})$ with $\mu$ denoting the chemical potential. 

Using the form factor in Eq.~\eqref{form_fac}, the non-interacting response function[$\chi^{\rm NI}({\bm q},\omega)$] is obtained using the following Hilbert transformation,
\begin{equation}
\label{chi0_hilbert}
\begin{split}
\chi^{\rm NI}=\int_{0}^{\infty}d\omega^{\prime}S^{\rm NI}({\bm q},\omega^{\prime})
\left[\frac{1}{\omega-\omega^{\prime}+i\eta}-\frac{1}{\omega+\omega^{\prime}+i\eta} \right].
\end{split}
\end{equation}
This is now used to calculate the dynamical dielectric function within RPA using Eq.~\eqref{epsiloneq}, and the plasmon modes can be obtained from Eq.~\eqref{root_epsilon}. We have used the background dielectric constant $\kappa=1$ in our calculation.

Experimentally, plasmon modes are generally probed by inelastic electron energy loss spectroscopy~\cite{EELS2014}. The collective density excitations  appear as peaks in the electron energy loss function spectrum (EELS), which is defined as 
% This can be matched with the loss function defined as,
\begin{equation}
{\cal E}_{loss}({\bm q},\omega) \approx  -{\rm Im}\left[\frac{1}{\epsilon^{\rm RPA}({\bm q},\omega)}\right].
\label{EELS}
\end{equation}

\begin{figure*}[t]
	\includegraphics[width = 0.9\linewidth]{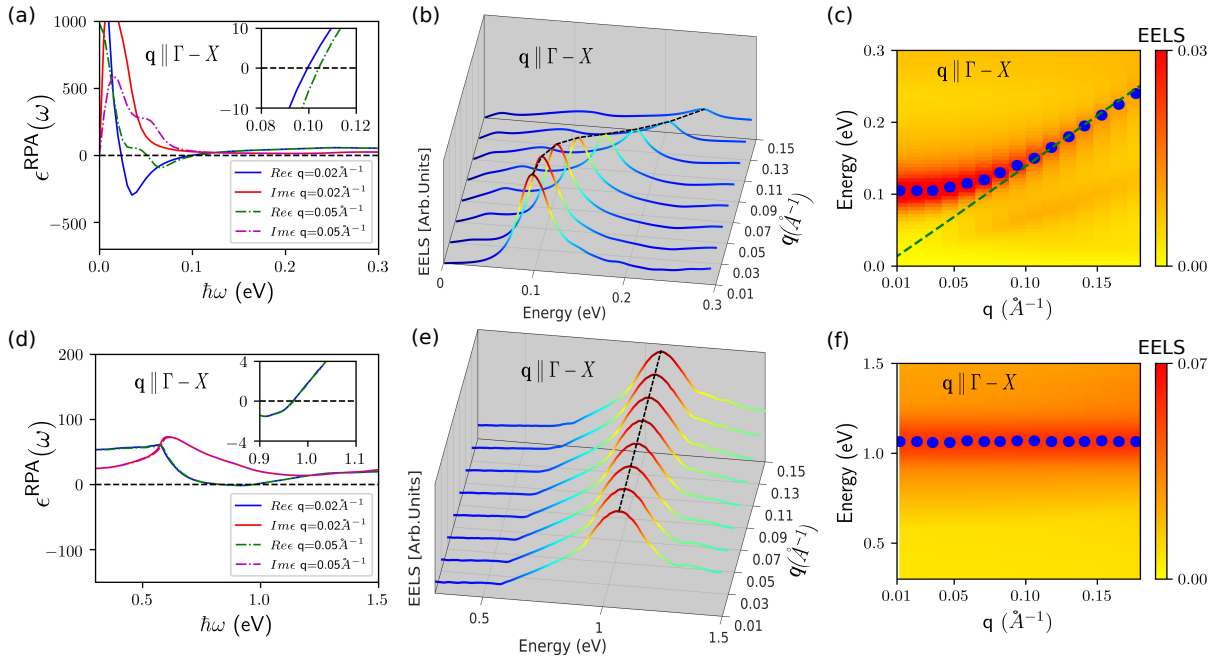}	
	\caption{(a) The dielectric function of CoSi for two different momentum transfers along the $\Gamma-X$ high symmetric direction. The magnified version of the $Re[\epsilon]$ is shown in the inset. (b) The EELS spectrum of CoSi for various momentum transfers. The black dashed line is a guide to the eye. (c) Colormap of the loss function spectrum in the $q-\omega$ plane. The green dashed line represents the boundary of the single-particle continuum. The blue dots represent the peak positions of the EELS spectra in panel (b). The panel (d), (e), and (f) are the same as (a), (b), and (c), respectively, but for a higher energy window. }
	\label{fig3}
\end{figure*}  

\subsection{Dielectric function and EELS of intrinsic ${\bf CoSi}$}
\begin{figure*}[t]
	\includegraphics[width = 1.0\linewidth]{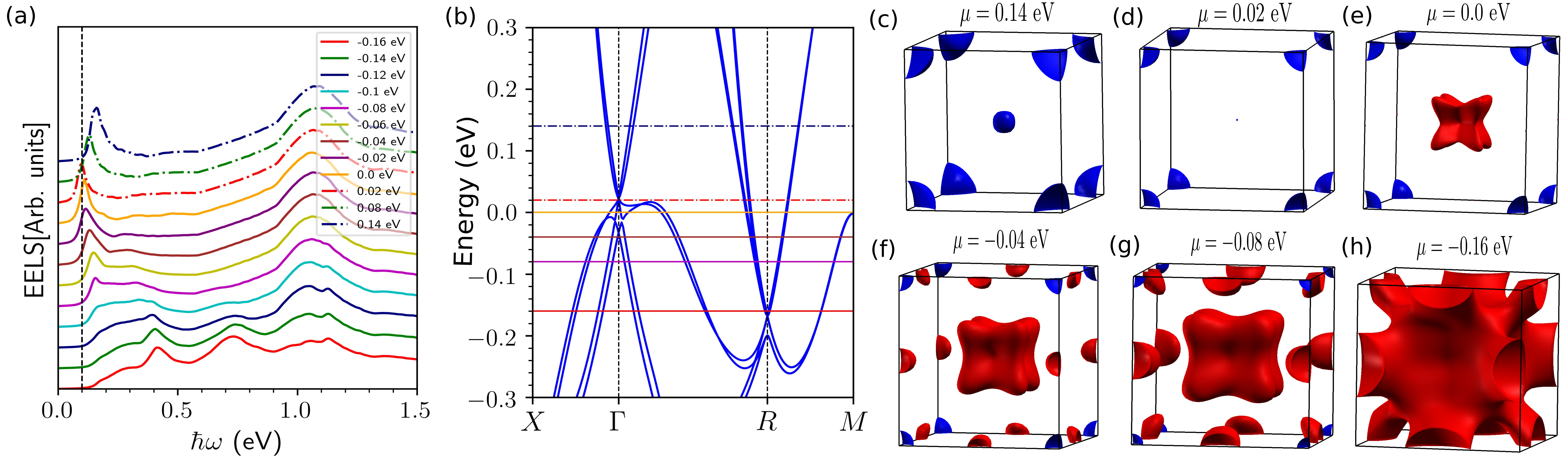}	
	\caption{(a) EELS spectrum of CoSi at small ${q}$ for different shifts in the Fermi energy. (b) Band structure of CoSi with various horizontal lines denoting the position of the Fermi energy. The color scheme of the horizontal lines in panel (b) are the same as the one used in panel (a). The corresponding Fermi surfaces are shown in panel (c)-(h). The red and blue colors denote the hole and the electron pockets, respectively.}
	\label{fig4}
\end{figure*}

We present the calculated dielectric function of CoSi, for two different momentum transfers along the $\Gamma-X$ high symmetry direction in Fig.~\ref{fig4}(a). The real part of the dielectric function (Re[$\epsilon$]) starts from a very high value and crosses the energy axis twice in the low energy window of 0-0.3 eV. Such a crossing where the ${\rm Re}[\epsilon]$ vanishes and the ${\rm Im}[\epsilon]$ is small is indicative of a plasmon mode with long lifetime. However, all such crossings do not specify a collective mode. To have a stable plasmon mode, the plasmon damping rate, $\Gamma_p \approx {\rm Im}[\epsilon (\omega\to \omega_p)]/(\partial_{\omega}{\rm Re}[\epsilon (\omega \to \omega_p)])$ should be a positive number~\cite{Agarwal2015}. In other words, in the vicinity of a crossing, the Im[$\epsilon$] and the derivative of the Re[$\epsilon$] should have the same sign, to have a plasmon mode. This condition is satisfied by the second crossing at $\sim 0.1$ eV where the imaginary part of the dielectric function is also small, implying a long-lived plasmon mode. The signature of this plasmon mode is clearly visible in the EELS peak shown in Fig.~\ref{fig3}(b). With an increase in the momentum transfer, the energy at which Re[$\epsilon$] vanish shifts towards higher energy [see inset of Fig.~\ref{fig3}(a)], and consequently, the plasmon mode is blue shifted. This is clearly visible in the EELS plot shown in Fig.~\ref{fig3}(b), where the peak position moves to higher energies with diminishing intensity for higher momentum transfers. 

The colormap of the EELS in the $q-\omega$ plane is shown in Fig.~\ref{fig3}(c) with the boundary of the intraband particle-hole continuum marked by the green dashed line. The plasmon dispersion remains outside the particle-hole continuum in the long wavelength limit. This indicates a long-lived plasmon mode that is not damped by any single-particle excitations, and it enters the particle hole continuum for a momentum transfer of $\sim$ 0.1 $\rm \AA^{-1}$.  The origin of this plasmon mode in CoSi with a plasmon gap of 0.1 eV can be understood from the electron pocket at the $R$  point arising from the double spin-1 excitation. The electron pocket at the $R$ point, has a chemical potential (measured from the band crossing point) of $~0.17$ eV. Now, considering the typical value of $\alpha_{\rm fine}=1$, we find $\Delta_p \approx 0.11$ eV, which is reasonably close to the value of $0.10$ eV, estimated from the {\it ab-initio} calculations. Thus, we find that the $0.10$ eV plasmon mode in CoSi has {\it dominant} contributions arising from the density fluctuations of the double spin-1 excitations. This conclusion is also supported by the investigation on doping presented in the next section. 

Interestingly, in a slightly higher energy window there is another energy (0.95 eV) where Re$[\epsilon]$ vanish [see Fig.~\ref{fig3}(d)], and the plasmon damping rate, $\Gamma =  {\rm Im}[\epsilon (\omega\to \omega_p)]/(\partial_{\omega}{\rm Re}[\epsilon (\omega \to \omega_p)]) >0$. This gives rise to another stable plasmon mode. However, in contrast to the previous mode, this mode is found to be dispersionless. The dispersionless nature of this mode is highlighted in the inset of Fig.~\ref{fig3}(d), where the `zero-crossing' point of ${\rm Re}[\epsilon]$ remains the same for two different momentum transfers. The EELS spectrum shown in Fig.~\ref{fig3}(e) further demonstrates the dispersionless nature of this plasmon mode, whose intensity also remains unaltered with increasing momentum transfer. In the EELS color map in the $q-\omega$ plane, this mode appears as a horizontal line [see Fig.~\ref{fig3}(f)]. This completely dispersionless plasmon mode arises from the interband correlations~\cite{PhysRevLett.119.266804,Jia_2020} (see Appendix~\ref{interband_plasmon} for details). 

Note that, although we have chosen the momentum transfer along the $\Gamma X$direction, the cubic symmetry ensures similar plasmonic excitations for the momentum transfer along the $\Gamma Y$ and $\Gamma Z$ direction as well. 

\section{EELS spectrum of doped ${\bf CoSi}$}   
\label{sec5}
We now study the effect of tuning the Fermi level in CoSi, which can be useful in unraveling the origin of the low energy plasmon modes. In the pristine CoSi, without any doping, there is an electron pocket centered around the $R$  point of the BZ and a hole pocket enclosing the $\Gamma$ point as shown in Fig.~\ref{fig4}(e). With an $n$-type doping, the size of this electron pocket increases while the hole pocket size shrinks, and it completely disappears for a doping with $\mu= 0.02$ eV[see Fig.~\ref{fig4}(d)]. Importantly, at this doping, the low-energy plasmon mode (with $\omega_p = 0.1$ eV) still exists. This confirms that the lowest energy plasmon mode in CoSi primarily originates from the collective charge density oscillations of the double spin-1 excitation around the $R$ point. On increasing the $n$-type doping further, and the plasmon energy is blue-shifted [see Fig.~\ref{fig4}(a)] due to the increasing size of the electron pocket and the increased carrier density of electrons.

On doing the $p$-doping, we find that the electron pocket at the $R$  point shrinks, the hole pocket around the $\Gamma$ point grows, and additionally, an extra hole pocket appears around the M point see Fig.~\ref{fig4}(f-h). On increasing the $p$-doping we find that the electron pocket at the $R$  point  shrinks in size and completely vanishes for $\mu = -0.16$ eV. Consequently, the lowest energy plasmon at ~0.1 eV, also vanishes. Remarkably, we find that the dispersionless plasmon peak around 1.1 eV is very robust and it remains largely unaltered for the full range of $n$- or $p$- doping with $\mu \in [-0.1, 0.14]$ eV.  

\section{Conclusions}
\label{sec6}
In summary, we demonstrate the existence of two long-lived plasmon modes in CoSi as an exemplar chiral multifold fermion semimetal using low energy modelling and density functional theory based first-principles calculations. Both plasmon modes are found to lie in the infrared region. The lowest energy mode primarily arises from the intraband collective density excitations of the double spin-1 fermion. It is highly dispersive with a plasmon gap of 0.1 eV in the long-wavelength limit. The higher energy mode with a plasmon gap of 1.1 eV is almost dispersionless, and it arises from the interband collective excitations. 
Such low-loss long-lived plasmons can have useful applications in optical and plasmonics devices in the infrared regime.

\section{Acknowledgments}
D.D. and A.A. acknowledge Science and Engineering Research Board (SERB) and Department of Science and Technology (DST), India for funding. 
The work at Northeastern University was supported by the Air Force Office of Scientific Research under award number FA9550-20-1-0322, and it benefited from the computational resources of Northeastern University's Advanced Scientific Computation Center (ASCC) and the Discovery Cluster.
\section{Appendix}

\appendix
\section{High-energy Plasmons in CoSi}
\begin{figure}[h!]
	\includegraphics[width = \columnwidth]{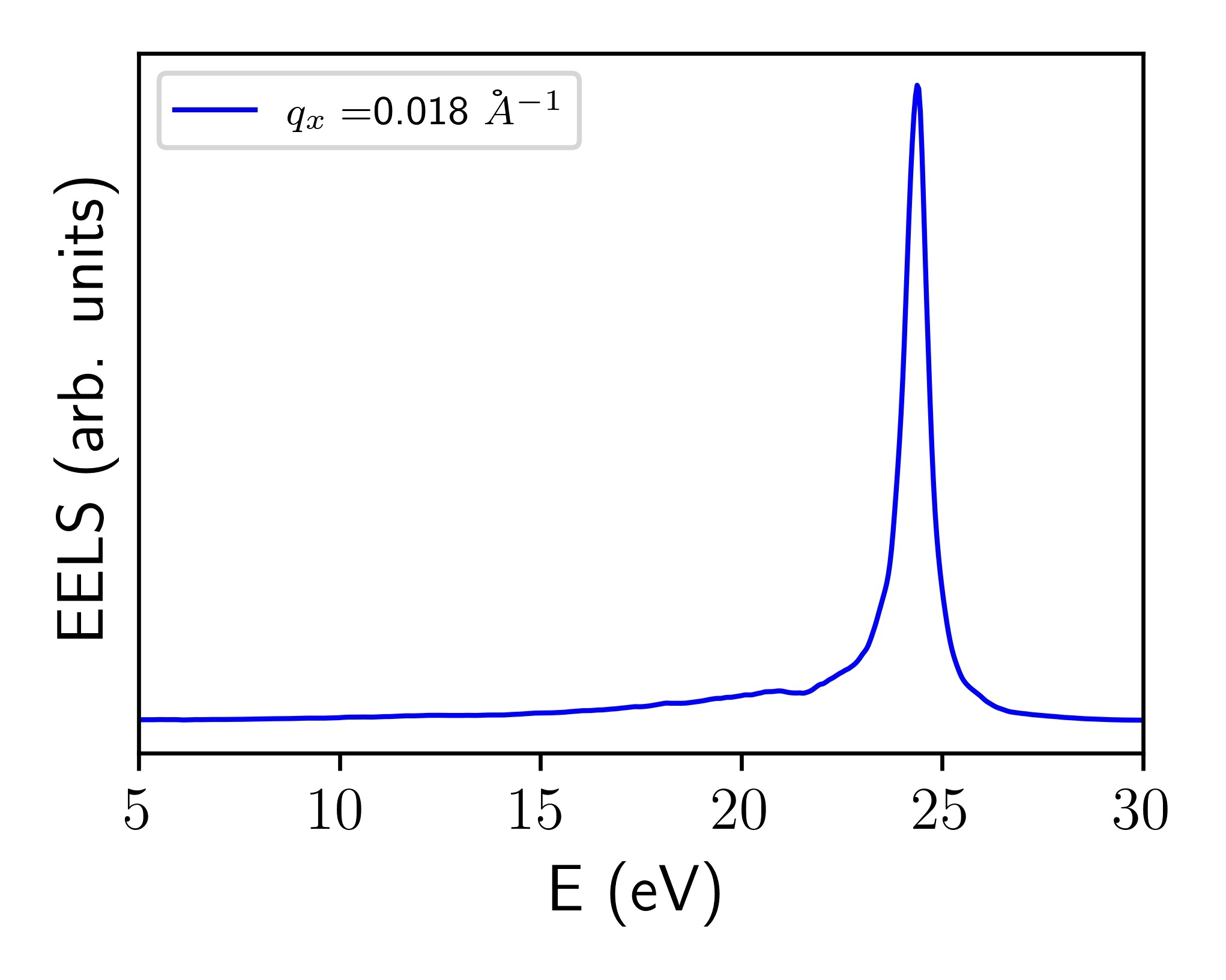}	
	\caption{ (a) EELS spectrum of CoSi calculated in a high energy window. The interband plasmon mode at 24 eV is evident.}
	\label{fig_high_en_plasmon}
\end{figure}

In the main text we focused on the low energy plasmons in CoSi. The high energy plasmon in the deep ultraviolet region($\sim$ 24 eV) is shown in Fig.~\ref{fig_high_en_plasmon}. This spectrum is calculated within the time dependent density functional theory(TDDFT) formalism using the RPA approximation as implemented in the GPAW package ~\cite{Enkovaara_2010,Thygesen2011}. The projected augmented-wave method and PBE exchange-correlation functionals are used for the ground state calculations.  

\section{Calculation of the density-density response function and the optical conductivity}\label{chi0_calculation}
Let us consider a translational invariant electronic system without long-range electron-electron interaction in the presence of an external perturbation ${V_{\rm{ext}}}(t)=\int d\bm{r}^{\prime}V_{\rm ext}(\bm{r}^{\prime},t)\hat{n}(\bm{r}^{\prime})$, where $\hat{n}(\bm{r})$ is the number density operator.  In the linear response regime, the induced density due to this external perturbation in the Fourier domain is given by~\cite{giuliani2005quantum},
\begin{equation}
n_{\rm{ind}}(\bm{q},\omega)=\chi^{\rm NI}(\bm{q},\omega)V_{\rm{ext}}(\bm{q},\omega)~.
\end{equation}
Here, $\chi^{\rm NI}{(\bm{q},\omega)}$ is the retarded non-interacting density-density response function or the Lindhard function.  Now the continuity equation $\partial_{t}n_{\rm{ind}}(\bm{r},t)+\nabla\cdot\bm{J}(\bm{r},t)=0$ in Fourier space becomes $-i\omega n_{\rm{ind}}(\bm{q},\omega)+i\bm{q}\cdot\bm{J}(\bm{q},\omega)=0$. Here, the electrical current is given by $\bm{J}(\bm{q},\omega)=\sigma(\bm{q},\omega)\bm{E}_{\rm ext}(\bm{q},\omega)$, in the presence of an external electric field $\bm{E}_{\rm ext}(\bm{q},\omega)=-i\bm{q}V_{\rm ext}(\bm{q},\omega)$. This connects the longitudinal optical conductivity and the dynamical non-interacting density-density response function via the relation,
\begin{eqnarray}
\sigma(\bm{q},\omega)&=&\frac{i\omega e^2}{q^2}\chi^{ \rm NI}{(\bm{q},\omega)}~.
\label{continuity}
\end{eqnarray}

The general expression of density-density response function in the spectral representation is given by~\cite{Adler_1962,giuliani2005quantum},
\begin{equation}
%	\begin{split}
	\chi^{\rm NI}=\frac{g}{V}\sum_{\bm{k},n,n^{\prime}}\frac{f_{n{\bm k}}-f_{n^{\prime}{\bm k}+{\bm q}}}{\hbar\omega+i0^{+}+\epsilon_{n{\bm k}} -\epsilon_{n^{\prime}{\bm k}+{\bm q}}} F_{nn^{\prime}}(\bm{k},\bm{k}+\bm{q})~.
%	\end{split}
	\label{chi0}
\end{equation}
The terms are described in main text [see section~\ref{Methodology}].
In the long-wavelength limit, $\bm{q}\to 0$, the plasmon gap ($\hbar\omega_p$) in three dimensional system can be calculated from the zero of the real part of dielectric function as [see Eq.~\eqref{root_epsilon1}],
\begin{equation}
1- V(q) {\rm Re}[\chi^{\rm NI}_{\rm interband} + \chi^{NI}_{\rm intraband}]=0~.
\label{epsilon}
\end{equation}

\subsection{Two-fold Weyl fermion}
The effective low energy Hamiltonian near a two-fold degenerate Weyl node is given by ${\cal H}=\hbar{v_F}\bm{\sigma}\cdot\bm{k}$. The corresponding eigen-spectrum is given by $\epsilon_{s{\bm k}}=s\hbar v_F{k}$, where ${s}=\pm{1}$ corresponds to the conduction and valence bands, respectively.
In the small $q$ limit, the intraband contribution of the density response function in Eq.~\eqref{chi0} can be approximately calculated by using a Taylor's expansion near the Fermi energy($\mu$). In the  $T\to 0$ limit, we have 
\small
\begin{eqnarray}
\chi^{\rm NI}_{\rm{intra}}(\bm{q}\to{0},\omega) &\simeq& -\frac{1}{(2\pi)^3}\int d^{3}\bm{k}~ \frac{ \bm{q}\cdot\nabla_{\bm{k}}\epsilon_{k}  }{\hbar\omega -\bm{q}\cdot\nabla_{\bm{k}}\epsilon_{k}}\cdot\frac{\partial f_{k}}{\partial \epsilon_k}\cdot 1  
\nn\\ & \simeq & \frac{1}{(2\pi)^3}\int d^{3}\bm{k} ~ \bm{q}\cdot(\nabla_{\bm{k}}\epsilon_{k})\cdot\frac{\delta(\mu-\epsilon_{k})}{\hbar\omega -\bm{q}\cdot\nabla_{\bm{k}}\epsilon_{k} }
\nn  \\ & \simeq & \frac{\mu^2}{2\pi^2\hbar^3v_F^3}\left[ -1 + \frac{\omega}{2v_F q}{\ln}\left(\frac{1+{v_F}q/\omega}{1-{v_F}q/\omega} \right) \right].
\nn \\
\end{eqnarray}
In the {\it dynamical} long wavelength limit ($q\to 0$ and $\omega\gg {q}v_F$), $\chi^{NI}(q,\omega)$ can be expanded as, 
\begin{equation}
{\rm Re}\left[\chi^{\rm NI}(\bm{q}\to{0},\omega)\right]= \frac{\mu^2}{6\pi^2\hbar^3v_F}\left( \frac{q}{\omega}\right)^2 + \mathcal{O}\left({\frac{q}{\omega}}\right)^4~.
\label{Weyl_intra}
\end{equation} 

Next, in the small $q$ limit, the interband density response function can be evaluated from the interband optical conductivity via Eq.~\eqref{continuity}. Though a direct calculation is straightforward, the optical conductivities are already calculated in the previous works, and this serves as an additional check of our calculations. In the limit of zero temperature, the real part of the interband optical conductivity for Weyl Hamiltonian having finite chemical potential $\mu$ at conduction band, is given by~\cite{deJuan2019}
\begin{equation}
	{\rm Re}\left[\sigma_{\rm{inter}}(\omega)\right]=\frac{\omega e^2}{24\pi\hbar{v_F}}\Theta(\hbar\omega-2\mu)~.
\end{equation}
Here $\Theta(x)$ is the Heaviside step function. By using the general Krammer's Kronig (KK) transformation, we can obtain the imaginary part of the optical conductivity as,
\begin{eqnarray}
{\rm{Im}}\left[\sigma_{\rm{inter}}(\omega)\right] &=& -\frac{1}{\pi}~P \int_{-\infty}^{\infty}d\omega^{\prime}\frac{{\rm Re}\left[ \sigma_{\rm inter}(\omega^{\prime})\right] }{ \omega^{\prime} -\omega}
\nn \\ &=& -\frac{\omega e^2}{24\pi^2\hbar{v_F}} {\ln}\left|\frac{\hbar^2\Lambda^2-\hbar^2\omega^2}{4\mu^2-\hbar^2\omega^2}\right|~.
\label{kk}
\end{eqnarray}  
Here, ${P}$ denotes the Cauchy principle value, and $\hbar\Lambda$ is the higher energy (or ultraviolet) cutoff for the linearly dispersing bands. In the $q \to 0$ limit, the interband response function upto the lowest of ${q}$ can be evaluated from Eq.~\eqref{continuity} as,
\begin{equation}
{\rm Re}\left[\chi_{\rm inter}^{\rm NI}{(\bm{q}\to{0},\omega)}\right]= -\frac{q^2}{24\pi^2\hbar{v_F}}{\ln}\left|\frac{\hbar^2\Lambda^2-\hbar^2\omega^2}{4\mu^2-\hbar^2\omega^2}\right|~.
\label{Weyl_inter}
\end{equation}
As the ultraviolet energy cutoff $\hbar\Lambda\gg\hbar\omega$, we can neglect this $\hbar\omega$ term in the numerator of Eq.~\eqref{Weyl_inter}. Finally, by using Eq.~\eqref{Weyl_intra}, Eq.~\eqref{Weyl_inter}, and Eq.~\eqref{epsilon}, we have obtained the long-wavelength plasmon gap for single Weyl node as given in Eq.~\eqref{weyl_plasmon}.

\subsection{Three-fold spin-1 excitation}
The band eigenvalues for threefold spin-1 Hamiltonian are given by $\epsilon_{sk}=s\hbar{v_F}{k}$, where $s=-1,~0,~1$. On account of the linearly dispersing bands, similar to that in a spin-1/2 Weyl node, the intra-band density response function for spin-1 excitation is also given by Eq.~\eqref{Weyl_intra}.
For spin-1 multifold Hamiltonian, only those vertical transitions which satisfy $\Delta{s}=s-s^{\prime}=\pm{1}$ are allowed. Therefore the real part of interband optical conductivity turns out to be~\cite{deJuan2019}
\begin{equation}
{\rm{Re}}\left[\sigma_{\rm{inter}}^{3f}(\omega)\right]= \frac{\omega e^2}{6\pi\hbar{v_F}}\Theta(\hbar\omega-\mu).
\end{equation} 
Using the KK transformation on Eq.~\eqref{kk}, we have
\begin{equation}
{\rm Im}\left[\sigma_{\rm{inter}}^{3f}(\omega)\right]= -\frac{\omega e^2}{6\pi^2\hbar{v_F}} \ln\left|\frac{\hbar^2\Lambda^2-\hbar^2\omega^2}{\mu^2-\hbar^2\omega^2}\right|~.
\end{equation}
Using this in Eq.~\eqref{continuity}, we obtain the interband non-interacting density-density response function in long-wavelength limit as
\begin{equation}
 	{\rm Re}\left[\chi_{\rm{inter}}^{\rm NI}{(\bm{q}\to 0,\omega)}\right]= -\frac{q^2}{6\pi^2\hbar{v_F}}{\ln}\left|\frac{\hbar^2\Lambda^2-\hbar^2\omega^2}{\mu^2-\hbar^2\omega^2}\right|~.
\end{equation}
The real part of dielectric function for spin-1 system can be evaluated from Eq.~\eqref{Reepsilon}, and it is given by 
\begin{equation}
     {\rm Re}\left[\epsilon^{\rm RPA}\right] =  1 + \frac{2\alpha_{ {\rm fine}}}{3\pi}\ln\left|\frac{\hbar^2\Lambda^2}{\mu^2-\hbar^2\omega^2} \right| - \frac{2\alpha_{ {\rm fine}}}{3\pi} \frac{\mu^2}{\omega^2}~.
\end{equation}
From the zeros of this equation, the plasmons gap is given by the self-consistent solution of the transcendental equation
\begin{equation}
	\hbar\omega_p=\Delta_p \left[ 1+ \frac{2\alpha_{{\rm fine}}}{3\pi}\ln\left|\frac{\hbar^2\Lambda^2}{\mu^2-\hbar^2\omega_p^2} \right| \right]^{-1/2}~.
\end{equation}
$\Delta_p$ and $\alpha_{{\rm fine}}$ are defined in main text.

\subsection{Fourfold double spin-1/2 fermion }
The fourfold double spin-1/2 Weyl fermion consists of two decoupled copies of spin-1/2 Weyl Hamiltonian as described in Eq.~\eqref{DoubleWeyl}. Due to double degeneracies, long-wavelength intraband response function up to the lowest $q$ order is given by
\begin{equation}
	{\rm Re}\left[\chi_{\rm{intra}}^{\rm NI}(\bm{q}\to{0},\omega)\right]= \frac{2\mu^2}{6\pi^2\hbar^3v_F}\left( \frac{q}{\omega}\right)^2~.
\end{equation}
Similarly, the interband response function in small ${q}$ limit is given by,
\begin{equation}
{\rm Re}\left[\chi_{\rm{inter}}^{NI}{(\textbf{q}\to{0},\omega)}\right]= -\frac{2q^2}{24\pi^2\hbar{v_F}}\rm{ln}\left|\frac{\hbar^2\Lambda^2-\hbar^2\omega^2}{4\mu^2-\hbar^2\omega^2}\right|~.
\end{equation}
Calculating the dielectric function, we find that the transcendental equation for the plasmon gap is given by 
\begin{equation}
\hbar\omega_p=\sqrt{2}\Delta_p \left[ 1+ \frac{2\alpha_{{\rm fine}}}{6\pi}\ln\left|\frac{\hbar^2\Lambda^2}{\mu^2-\hbar^2\omega_p^2} \right| \right]^{-1/2}.
\end{equation}

\subsection{Fourfold spin-3/2 fermion}
The band eigenvalues for spin-3/2 Hamiltonian are given by $\epsilon_{sk}=2s\hbar{v_F}k$, with ${s}=-\frac{3}{2},-\frac{1}{2},~\frac{1}{2},~\frac{3}{2}$.  Now for finite chemical potential $\mu$ in conduction bands, only two bands with dispersion $\epsilon_{1}=\hbar v_F{k}$ and $\epsilon_{2}=3\hbar v_F{k}$ will contribute to intra-band response function. So, intraband density response in small ${q}$ limit can be approximated as,
\begin{eqnarray}
\chi_{\rm{intra}}^{\rm NI}(\bm{q}, \omega) \simeq  -\frac{1}{(2\pi)^3}\sum_{s^{\prime}=1,2}\int d^{3}\bm{k}~ \frac{ \bm{q}\cdot\nabla_{\bm{k}}\epsilon_{s^{\prime}}  }{\hbar\omega -\bm{q}\cdot\nabla_{\bm{k}}\epsilon_{s^{\prime}}}\cdot\frac{\partial f_{s^{\prime}k}}{\partial \epsilon_{s^{\prime}}} ~.\nn \\
%\nn \\ &\simeq& \frac{1}{(2\pi)^3}\sum_{s=1,2}\int d^{3}\textbf{k}~\frac{ \textbf{q}\cdot\nabla_{\bm{k}}\epsilon_{sk}  }{\hbar\omega -\textbf{q}\cdot\nabla_{\bm{k}}\epsilon_{sk}}\cdot\delta(\mu-\epsilon_{sk})~. \nn \\
\label{spin3/2_intra1}
\end{eqnarray}
By solving Eq.~\eqref{spin3/2_intra1}in the {\it dynamical} long wavelength limit, we obtain the intraband density-density response function upto lowest order of $q$ to be,
\begin{eqnarray}
	{\rm{Re}}\left[\chi_{\rm intra}^{\rm NI}(\bm{q}\to{0},\omega) \right] &\simeq & \frac{4\mu^2}{18\pi^2\hbar^3v_F}\left(\frac{q}{\omega}\right)^2.
	\label{spin3/2_intra_2}
\end{eqnarray}

The optical conductivity for this spin-3/2 system can be calculated analytically for zero temperature, and its real part is given by~\cite{deJuan2019}
\begin{eqnarray}
{\rm{Re}}\left[\sigma_{\rm{inter}}^{4f}(\omega)\right]=\frac{\omega e^2}{8\pi\hbar{v_F}}\left[ \frac{1}{3}\Theta(\hbar\omega-2\mu) + \Theta(\hbar\omega-\frac{2}{3}\mu)  \right]~. \nn \\
\end{eqnarray}
The imaginary part of the optical conductivity can be obtained by using KK transformation as described in Eq.~\eqref{kk} and it is given by,
\begin{eqnarray}
{\rm{Im}}\left[\sigma_{\rm{inter}}^{4f}(\omega)\right] = & &-\frac{\omega e^2}{24\pi^2\hbar{v_F}} \left( 4~\ln\left|\frac{\hbar^2\Lambda^2-\hbar^2\omega^2}{4\mu^2-\hbar^2\omega^2}\right| \right. 
\nn \\
& &\left. +3~\ln\left|\frac{36\mu^2-9\hbar^2\omega^2}{4\mu^2-9\hbar^2\omega^2}\right| \right)~.
\nn \\
\end{eqnarray} 
Therefore, in small $q$ limit, interband density response function can be evaluated from Eq.~\eqref{continuity} as,
\begin{eqnarray}
{\rm Re}\left[\chi_{\rm{inter}}^{0}{(\bm{q}\to{0},\omega)}\right]= & & -\frac{q^2}{24\pi^2\hbar{v_F}} \left( 4~\ln\left|\frac{\hbar^2\Lambda^2-\hbar^2\omega^2}{4\mu^2-\hbar^2\omega^2}\right| \right. 
\nn \\
& &\left. +3~\ln\left|\frac{36\mu^2-9\hbar^2\omega^2}{4\mu^2-9\hbar^2\omega^2}\right| \right) ~.
\nn \\
\label{spin3/2_inter}
\end{eqnarray} 
So, by substituting Eq.~\eqref{spin3/2_intra_2} and Eq.~\eqref{spin3/2_inter} in Eq.~\eqref{epsilon}, we can obtained the transendental equation for the plasmon gap $\hbar\omega_p$ to be,

\begin{eqnarray}
\hbar\omega_p= & &\frac{2}{\sqrt{3}}\Delta_p \left[ 1 + \frac{\alpha_{{\rm fine}}}{6\pi}\left( 4~\ln\left|\frac{\hbar^2\Lambda^2}{4\mu^2-\hbar^2\omega_p^2}\right| \right.\right. \nn 
\\ 
& &\left.\left. +3~\ln\left|\frac{36\mu^2-9\hbar^2\omega_p^2}{4\mu^2-9\hbar^2\omega_p^2}\right|   \right) \right]^{-1/2}~.
\end{eqnarray}

\subsection{Sixfold double spin-1 excitation}
The effective low energy Hamiltonian for sixfold double spin-1 excitation consists of two decouple copies of spin-1 Hamiltonian as described in Eq.~\eqref{Doublespin1_H}. Therefore, in small $q$ limit, the non-interacting density response function is just twice that of the spin-1 case
\begin{equation}
{\rm Re}\left[\chi^{\rm NI}_{\rm{intra}}(\bm{q}\to{0},\omega)\right]= \frac{2\mu^2}{6\pi^2\hbar^3v_F}\left( \frac{q}{\omega}\right)^2~.
\end{equation} 
Similarly, interband density-density response function is given by 
\begin{equation}
{\rm Re}\left[\chi_{\rm{inter}}^{\rm NI}{(\bm{q}\to 0,\omega)}\right]= -\frac{2q^2}{6\pi^2\hbar{v_F}}\ln\left|\frac{\hbar^2\Lambda^2-\hbar^2\omega^2}{\mu^2-\hbar^2\omega^2}\right|~.
\end{equation}
Therefore the long-wavelength plasmon gap $\hbar\omega_p$ is specified by the transendental equation,
\begin{equation}
	\hbar\omega_p=\sqrt{2}\Delta_p \left[ 1+ \frac{4\alpha_{{\rm fine}}}{3\pi}\ln\left|\frac{\hbar^2\Lambda^2}{\mu^2-\hbar^2\omega_p^2} \right| \right]^{-1/2}.
\end{equation}

\section{Origin of the interband plasmon at ~1.1 eV}
%\label{interband_plasmon}
\begin{figure}[t!]
	\includegraphics[width = \linewidth]{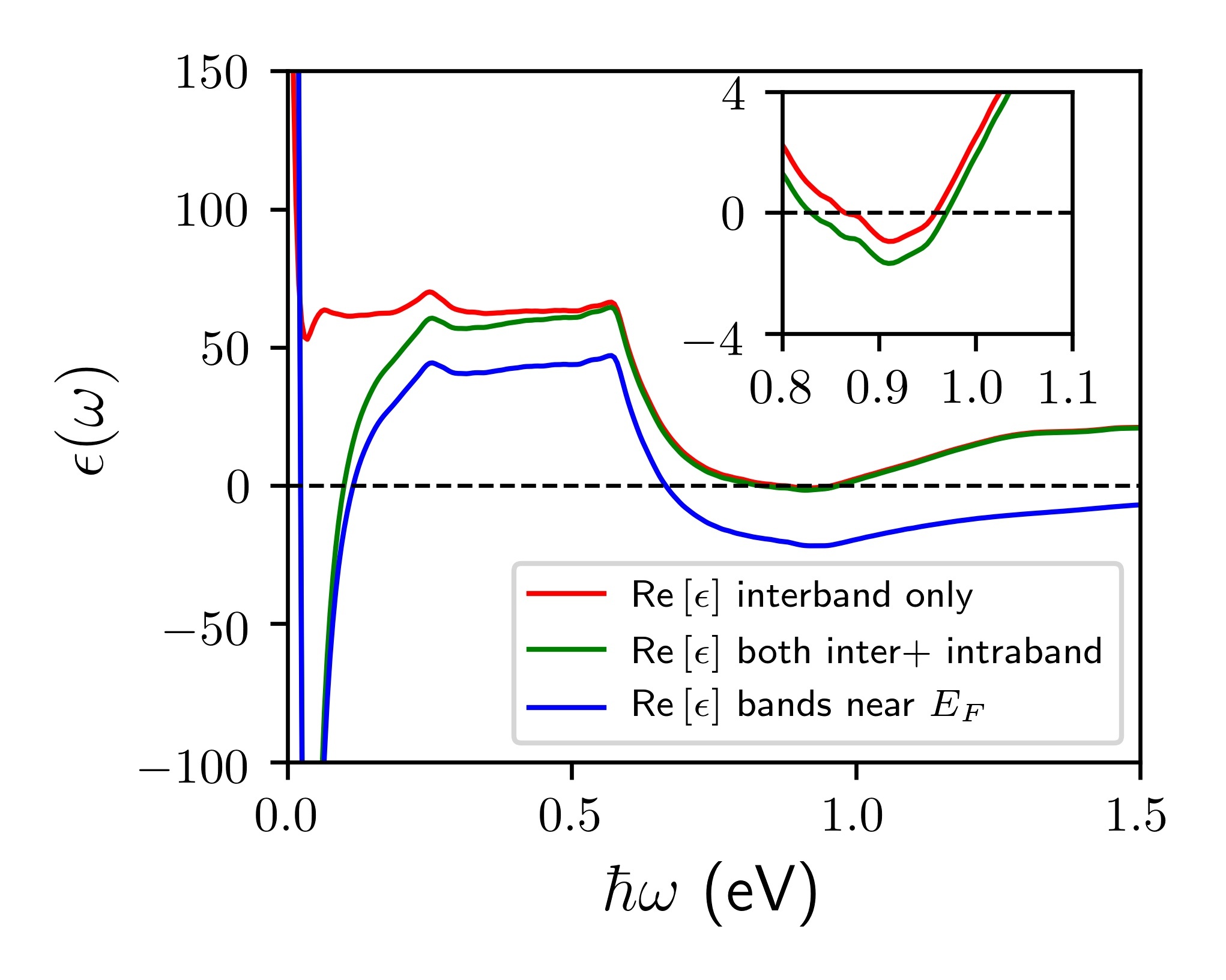}	
	\caption{ Real part of dielectric funtion at small $q$ ($q=$0.02 $\rm \AA^{-1}$), calculated for three different cases: (i) by including only interband terms between all the bands, (ii) both intra and interband correlations of all bands, (iii) both intra and interband correlations between the eight bands near Fermi level as shown in Fig.~\ref{fig1}(d). The inset plot shows the zoomed in version. }
	\label{interband_plasmon}
\end{figure} 
We have shown the existence of a dispersionless plasmon mode at $\sim$ 1.1 eV in CoSi in Fig.~\ref{fig3}, and argue that this mode is an `interband' plasmon mode [see Fig.~\ref{fig3}.(e) and (f)].  The term interband highlights that the interband correlation ($n\ne{n^{\prime}}$) in the density-density response function calculation are responsible for this mode~\cite{PhysRevLett.119.266804}. To confirm this explicitly, we show the real part of the dielectric function in Fig.~\ref{interband_plasmon}, for three different cases: (i) by considering only interband correlations of all bands (red line), (ii) both intraband and interband correlations of all bands (green line), and (iii) both intra and interband contributions of eight bands near Fermi level with SOC (blue line). The root of dielectric function at $\sim$ 0.1 eV is not present when we consider the interband correlation only (case-(i)), which confirms that this mode is an intraband plasmon mode. However, in case(i) the other root at $\sim$ 0.95 eV is still present. %In case-(ii), we have both roots at $\sim$0.1 eV and $\sim$0.95 eV.  
Case-(iii), which considers only intra and interband correlations between the eight bands near the Fermi level [see Fig.~\ref{fig1}(d)] involved in producing the multifold chiral fermions, misses the dispersionless plasmon mode in vicinity of 1.1 eV, while capturing the mode at 0.1 eV. This establishes that the 0.1 eV mode is an intraband plasmon mode and the 1.1 eV mode is an interband plasmon mode, which is not related to the bands near the Fermi energy. The inter-band plasmon mode arises only when we include the lower valence bands in our calculation. Note that, the plasmon peak in the EELS spectra ($\sim$1.1 eV) is shifted slightly compared to the root of the Re[$\epsilon$] ($\sim$0.95 eV) due to the presence of finite Im[$\epsilon$].

\bibliography{CoSi_v1}
\end{document}